\documentclass[pra,twocolumn,showpacs,floatfix]{revtex4-2}
\usepackage{graphicx}
\usepackage{dcolumn}
\usepackage{bm}
\usepackage{color}
\usepackage{amsmath}
\usepackage{amsfonts}
\usepackage{amssymb}
\usepackage{subcaption}
\usepackage[utf8]{inputenc}
\usepackage{caption}
\usepackage{overpic}
\usepackage{xcolor}
\usepackage{hyperref}
\usepackage{float}

\hypersetup{colorlinks=true, linkcolor=blue, citecolor=blue, urlcolor=blue, filecolor=blue}

\begin{document}

\title{Macroscopic quantum states, quantum phase transition for $N$ \\
three-level atoms in an optical cavity\\
----- Gauge principle and non-Hermitian Hamiltonian}
\author{Ni Liu}
\thanks{Contact author: liuni2011520@sxu.edu.cn}
\author{Xinyu Jia}
\author{J.-Q. Liang}
\thanks{Contact author: jqliang@sxu.edu.cn}

\begin{abstract}
We study in this paper the quantum phase transition (QPT) from normal phase
(NP) to superradiant phase (SP) for $N$ three-level atoms in a single-mode
optical cavity for both Hermitian and non-Hermitian Hamiltonians, where the $%
\Xi $-type three-level atom is described by spin-$1$ pseudo-spin operators.
The long standing gauge-choice ambiguity of $\mathbf{A\cdot p}$ and $\mathbf{%
d\cdot E}$ called respectively the Coulomb and dipole gauges is resolved by
the time-dependent gauge transformation on the Schr\"{o}dinger equation.
Both $\mathbf{A\cdot p}$ and $\mathbf{d\cdot E}$ interactions are included
in the unified gauge, which is truly gauge equivalent to the minimum
coupling principle. The Coulomb and dipole interactions are just the special
cases of unified gauge. Remarkably three interactions lead to the same
results under the resonant condition of field-atom frequencies, while
significant difference appears in red and blue detunings. The QPT is
analyzed in terms of spin-coherent-state variational method, which indicates
the abrupt changes of energy spectrum, average photon number as well as the
atomic population at the critical point of interaction constant. Crucially,
we reveal the sensitive dependence on the initial optical-phase, which is
particularly useful to test the validity of three gauges experimentally. The
non-Hermitian atom-field interaction results in the exceptional point (EP),
beyond which the semiclassical energy function becomes complex. However the
energy spectrum of variational ground state is real in the absence of EP,
and does not become complex. The superradiant state is unstable due to the
non-Hermitian interaction induced photon-number loss. Thus only the NP
exists in the non-Hermitian Dicke Model Hamiltonian.
\end{abstract}

\maketitle

\affiliation{Institute of Theoretical Physics, State Key Laboratory of Quantum Optics and
Quantum Optics Devices, Shanxi University, Taiyuan 030006, Shanxi, China}

\section{INTR\textbf{ODUCTION}}

Following the development of quantum computers \cite%
{Arute2019,Zhong2020,PhysRevLett.127.180502,Madsen2022}, the quantum
manipulations based on cavity quantum electrodynamics grow incredibly fast 
\cite{Stassi2020} in the ultrastrong-coupling regime.\textbf{\ }As a result
the field-atom coupling coefficient is comparable with the atom frequency,
so that the quantum phase transition (QPT) becomes realistic in experiments 
\cite{Baumann2010,Niemczyk2010}. The ultrastrong or even superstrong
coupling regimes have been realized experimentally in superconducting
circuits \cite{Niemczyk2010,Yoshihara2017,PhysRevApplied.13.054063} and
various other platforms \cite{FriskKockum2019}.

The QPT from normal phase (NP) to superradiant phase \ (SP) in Dicke model
(DM), which describes $N$ identical two-level atoms in a single-mode optical
cavity, has been extensively investigated theoretically and experimentally 
\cite%
{FriskKockum2019,PhysRev.93.99,PhysRevLett.36.1035,Scheibner2007,PhysRevLett.30.309,Bohnet2012,Kim2025,PhysRevResearch.6.043128,PhysRevA.90.023622,PhysRevA.93.033630,Lian_2012}%
. Multi-level atomic models, which are closer to realistic atomic structures,%
\textbf{\ }have attracted significant attention \cite%
{PhysRevLett.133.113601,PhysRevResearch.7.013288} in recent years\textbf{. }%
The energy spectra of three-level atoms have been studied for $\Lambda $$-$ 
\cite{PhysRevA.105.053702}, $V-$ \cite{PhysRevA.107.033711}, and $\Xi -$type 
\cite{PhysRevA.108.033706} of level configurations.\textbf{\ }The
three-level configuration of atoms is related to an important class of
quantum phenomena, including electromagnetically induced transparency \cite%
{PhysRevLett.66.2593,RevModPhys.77.633}, lasing without inversion \cite%
{PhysRevLett.62.2813,JMompart2000}, and quantum batteries \cite%
{Dou2020,Dou2021,PhysRevE.101.062114} as well.

\textbf{\ }The extension from the two-level to three-level DM naturally
provides a new perspective of atom-photon interactions\cite{Sung1979}.%
\textbf{\ }The three-level DM exhibits rich physics, including subradiance 
\cite%
{Crubellier1985,Crubellier1986,PhysRevA.79.053622,PhysRevLett.121.173602},
SP transitions \cite%
{PhysRevA.84.053856,PhysRevA.86.063822,PhysRevA.87.023813,PhysRevA.87.023805}%
, time crystals \cite{PhysRevA.104.063705,PhysRevLett.127.253601}, and
chiral molecule detection\cite{PhysRevResearch.4.013100}.

The electron-photon interaction seems formally dependent on the gauge choice 
\cite{Babiker1983}, which remains a long standing ambiguity, since
measurable results must be gauge independent.\textbf{\ }Gauge invariance is
a general requirement for the fundamental interactions \cite%
{Aitchison1989,maggiore2005modern} in modern quantum field theory \cite%
{Aitchison1989,maggiore2005modern,Wilczek2005,Wilczek2013}. Although the
Hamiltonian of minimal coupling is gauge invariant, the approximate
Hamiltonians may lead to different predictions for the light-matter
interaction derived in different gauges \cite%
{PhysRev.85.259,PhysRevA.36.2763,PhysRevA.3.1242,PhysRevB.24.2009,PhysRevLett.87.087402}%
.\textbf{\ }Moreover, the convergence of the two-photon transition rate
depends on the choice of the gauge significantly \cite{PhysRevLett.39.1070}.
The gauge principle is violated by the quantum Rabi model with the dipolar
coupling between a two-level atom and a quantized electromagnetic field \cite%
{PhysRevA.97.043820,PhysRevA.98.053819,Stokes2019}.\textbf{\ }This model has
been used to describe various quantum systems and physical processes under
different interaction regimes \cite{FriskKockum2019,Leonardi1986}.\textbf{\ }%
The predictions based on the Rabi model drastically depend on the chosen
gauge in the ultrastrong coupling regime.\textbf{\ }This gauge dependence is
also an attribute in the finite-level truncation of the matter system.

The gauge ambiguities, which result from two fundamentally equivalent
Hamiltonians, namely the minimal coupling and electric dipole Hamiltonians 
\cite{Rzaewski2004} respectively with $\mathbf{A\cdot p}$\textbf{\ }\ and $%
\mathbf{d\cdot E}$ interactions \cite{loudon2000quantum,DiStefano2019}. In
the Rabi Hamiltonian, which is derived \cite{Stokes2024} from the minimal
coupling, it is demonstrated that the failure of gauge invariance comes from
the improper\textbf{\ }two-level truncation of the atomic Hilbert space \cite%
{milonni2019introduction}.\textbf{\ }The gauge ambiguities have been
resolved recently \cite{PhysRevLett.125.123602} with a proper way of
applying two-level truncation, from which the gauge-invariant Rabi
Hamiltonian is obtained.

We resolve the gauge ambiguity in this work from an alternative viewpoint.
Since the Hamiltonian is explicitly time dependent. The time-dependent gauge
transformation has to be applied on the time-dependent Schr\H{o}dinger
equation. The gauge equivalent Rabi Hamiltonian to the minimum coupling
includes both the $\mathbf{A\cdot p}$\ \ and $\mathbf{d\cdot E}$ \
interactions, which we called the unified gauge. The QPT for $N$ three-level
atoms in a single-mode optical cavity is analyzed in term of the
spin-coherent-state (SCS) variational method \cite%
{PhysRevA.90.023622,PhysRevA.93.033630,Lian_2012,PhysRevA.110.063320}
respectively in Coulomb, dipole and unified gauges as comparison. The novel
phase effect of the optical field is incorporated in the energy function of
unified gauge Hamiltonian, which reduces to the Coulomb and dipole gauges in
the special phase value.

Non-Hermitian light-matter interactions have attracted considerable
attention in both the semiclassical regime \cite%
{wang2023non,Meng2024,ElGanainy2019} and the quantum regime \cite%
{Lodahl2017,PhysRevX.13.031009,Wang2022,PhysRevLett.131.160801}, owing to
their relevance to realistic physical scenarios of experiments \cite%
{PhysRevA.102.032202}. Unconventional features of the SP transition have
been revealed by incorporating non-Hermitian effects such as non-reciprocal
couplings\cite{PhysRevLett.131.113602} or complex potentials\cite%
{PhysRevA.108.053712,Liu2022} into the DM. In the cavity dynamic system the
non-Hermitian Hamiltonian exists naturally since the dipole matrix element $%
x_{ij}$ is a complex number with equal order of the real and imaginary parts 
\cite{peng1998introduction}. The real value of $x_{ij}$ results in the
non-Hermitian atom-field interaction. An exceptional point (EP) indeed
emerges in the semiclassical energy function. However, the energy spectrum
of variational ground state is real, with no complex value.

\section{\textbf{\ Hamiltonian of three-level atom in an optical cavity and
the gauge principle }}

\textbf{\ }

The Hamiltonian for an electron in an atom can be written as%
\begin{equation*}
H_{a}=\frac{\mathbf{p}^{2}}{2m}+eV,
\end{equation*}%
with three orthogonal eigenvectors denoted by $H_{a}|u_{i}\rangle
=E_{i}|u_{i}\rangle $, where $i=1,2,3$. The second-quantization Hamiltonian
can be formally obtained in terms of the completeness projection operator 
\begin{equation*}
\widehat{P}=\sum_{i=1,2,3}|u_{i}\rangle \left\langle u_{i}\right\vert ,
\end{equation*}%
such as 
\begin{equation*}
\widehat{H}_{a}=\widehat{P}H_{a}\widehat{P}=\sum_{i=1,2,3}E_{i}|u_{i}\rangle
\left\langle u_{i}\right\vert ,
\end{equation*}%
where the operator with hat denotes the second-quantization operator.
According to the minimum coupling the Hamiltonian of electron in an optical
cavity is 
\begin{equation}
H\left( t\right) =\frac{[\mathbf{p}-e\mathbf{A}\left( t\right) ]^{2}}{2m}+eV,
\label{2}
\end{equation}%
in the unit convention $c=\hbar =1$ with $\mathbf{A}\left( t\right) $ being
the time-dependent vector potential. The canonical momentum $\mathbf{p}=-i%
\mathbf{\triangledown }$ is as usual.

\subsection{\emph{Coulomb gauge}\textbf{\ }}

The conventional Hamiltonian called Coulomb gauge with $\nabla \cdot \mathbf{%
A}=0$ is derived as following 
\begin{equation}
\widehat{H}\left( t\right) =\widehat{P}H\widehat{P}%
=\sum_{i=1,2,3}E_{i}|u_{i}\rangle \left\langle u_{i}\right\vert +\frac{%
\widehat{A}^{2}}{2m}+\widehat{H}_{i},  \label{3}
\end{equation}%
in which the interaction part of Hamiltonian is%
\begin{equation*}
\widehat{H}_{i}\left( t\right) =-\sum_{i,j}|u_{i}\rangle \left\langle
u_{i}\right\vert \frac{\widehat{A}\cdot \mathbf{p}}{m}|u_{j}\rangle
\left\langle u_{j}\right\vert .
\end{equation*}

Using the relation 
\begin{equation}
\frac{\mathbf{p}}{m}=\overset{\cdot }{\mathbf{x}}=i\left[ H_{a},\mathbf{x}%
\right]  \label{1}
\end{equation}%
the interaction part is obtained as%
\begin{equation}
\widehat{H}_{i}\left( t\right) =i\widehat{A}\cdot \sum_{i\neq j}\mathbf{x}%
_{ij}\left( E_{i}-E_{j}\right) |u_{i}\rangle \left\langle u_{j}\right\vert ,
\label{4}
\end{equation}%
where the quantized vector potential is%
\begin{equation*}
\widehat{A}\left( t\right) =\xi \mathbf{e}\left( \widehat{a}e^{-i\omega t}+%
\widehat{a}^{\dag }e^{i\omega t}\right) ,
\end{equation*}%
in the long-wave-length approximation with $\xi =\sqrt{\frac{1}{2\omega
\epsilon _{0}}}$, and $\mathbf{e}$ being polarization unit vector. $\mathbf{x%
}_{ij}=\left\langle u_{i}\right\vert \mathbf{x}|u_{j}\rangle $ denotes the
matrix elements of the electron position vector. We assume $\mathbf{x}_{12}=%
\mathbf{x}_{23}$, and $\mathbf{x}_{13}=0$, which means that the cavity field
does not generate the transition between levels $1$ and $3$. The dipole
matrix element $\mathbf{x}_{12}$ is usually a complex number \cite%
{peng1998introduction} with actually equal order of the real and imaginary
parts \cite{peng1998introduction}. The interaction Hamiltonian in Eq. (\ref%
{4}) becomes%
\begin{eqnarray}
\widehat{H}_{i}\left( t\right) &=&i\xi \mathbf{e}\left( \widehat{a}%
e^{-i\omega t}+\widehat{a}^{\dag }e^{i\omega t}\right)  \notag \\
&&\cdot \lbrack \left( E_{3}-E_{2}\right) (\mathbf{x}_{32}|u_{3}\rangle
\left\langle u_{2}\right\vert -\mathbf{x}_{23}|u_{2}\rangle \left\langle
u_{3}\right\vert )  \label{5} \\
&&+\left( E_{2}-E_{1}\right) (\mathbf{x}_{21}|u_{2}\rangle \left\langle
u_{1}\right\vert -\mathbf{x}_{12}|u_{1}\rangle \left\langle u_{2}\right\vert
)].  \notag
\end{eqnarray}

We adopt the dimensionless coupling constant $G=\xi \beta $ with 
\begin{equation}
i\beta =\mathbf{e\cdot x}_{12},\quad -i\beta =\mathbf{e\cdot x}_{21},
\label{5-1}
\end{equation}%
in which the dipole matrix element $\mathbf{x}_{12}$ is regarded as the
imaginary number in consistent with the $J-C$ model. For the equal
level-space $\Xi $-type atom by setting $E_{2}=0,E_{3,1}=\pm \Omega $
respectively, the Hamiltonian becomes%
\begin{eqnarray}
\widehat{H}\left( t\right) &=&\Omega \left( |u_{3}\rangle \left\langle
u_{3}\right\vert -|u_{1}\rangle \left\langle u_{1}\right\vert \right)  \notag
\\
&&+G\Omega \left( \widehat{a}e^{-i\omega t}+\widehat{a}^{\dag }e^{i\omega
t}\right)  \label{6} \\
&&\cdot \left( |u_{3}\rangle \left\langle u_{2}\right\vert +|u_{2}\rangle
\left\langle u_{1}\right\vert +|u_{2}\rangle \left\langle u_{3}\right\vert
+|u_{1}\rangle \left\langle u_{2}\right\vert \right) ,  \notag
\end{eqnarray}%
in which the quadratic field term $\frac{\widehat{A}^{2}}{2m}$ is neglected
in the weak field approximation. The Hamiltonian can be represented in terms
of spin$-1$ operator, 
\begin{equation}
\widehat{H}_{c}=\Omega \widehat{s}_{z}+\frac{G\Omega }{\sqrt{2}}\left( 
\widehat{a}e^{-i\omega t}+\widehat{a}^{\dag }e^{i\omega t}\right) \left( 
\widehat{s}_{+}+\widehat{s}_{-}\right) ,  \label{7}
\end{equation}%
with 
\begin{eqnarray*}
\widehat{s}_{z} &=&[|u_{3}\rangle \left\langle u_{3}\right\vert
-|u_{1}\rangle \left\langle u_{1}\right\vert ], \\
\widehat{s}_{+} &=&\sqrt{2}\left( |u_{3}\rangle \left\langle
u_{2}\right\vert +|u_{2}\rangle \left\langle u_{1}\right\vert \right) ,\quad 
\widehat{s}_{-}=\widehat{s}_{+}^{\dag },
\end{eqnarray*}%
It is easy to check that the spin commutation relations are satisfied,%
\begin{equation*}
\left[ \widehat{s}_{z},\smallskip \widehat{s}_{\pm }\right] =\pm \widehat{s}%
_{\pm },\smallskip \quad \left[ \widehat{s}_{+},\smallskip \widehat{s}_{-}%
\right] =2\widehat{s}_{z},
\end{equation*}%
and 
\begin{equation*}
\left[ \widehat{s}_{x},\smallskip \widehat{s}_{y}\right] =i\widehat{s}_{z},
\end{equation*}%
with%
\begin{equation*}
\widehat{s}_{x}=\frac{\widehat{s}_{+}+\widehat{s}_{-}}{2},\quad \widehat{s}%
_{y}=\frac{\widehat{s}_{+}-\widehat{s}_{-}}{2i}.
\end{equation*}

We apply the time-dependent gauge transformation $|\psi ^{\prime }\rangle =%
\widehat{U}\left( t\right) |\psi \rangle $%
\begin{equation}
i\frac{\partial }{\partial t}|\psi ^{\prime }\rangle =\widehat{H}^{\prime
}|\psi ^{\prime }\rangle ,  \label{7-1}
\end{equation}%
where%
\begin{equation}
\widehat{H}^{\prime }=\widehat{U}\widehat{H}\widehat{U}^{\dag }-i\widehat{U}%
\frac{\partial }{\partial t}\widehat{U}^{\dag }.  \label{7-2}
\end{equation}%
Using unitary transformation operator 
\begin{equation}
\widehat{U}=e^{-i\omega t\widehat{a}^{\dag }\widehat{a}},  \label{7-3}
\end{equation}%
the final Hamiltonian of Coulomb gauge is time-independent%
\begin{equation}
\widehat{H}_{c}=\widehat{H}^{\prime }=\widehat{H}_{0}+\widehat{H}_{i}^{c},
\label{8}
\end{equation}%
where%
\begin{equation}
\widehat{H}_{0}=\omega \widehat{a}^{\dag }\widehat{a}+\Omega \widehat{s}_{z}
\label{9}
\end{equation}%
and%
\begin{equation}
\widehat{H}_{i}^{c}=\frac{G\Omega }{\sqrt{2}}\left( \widehat{a}+\widehat{a}%
^{\dag }\right) \left( \widehat{s}_{+}+\widehat{s}_{-}\right) ,  \label{10}
\end{equation}%
is the Coulomb gauge interaction.

\subsection{\protect\bigskip \emph{Dipole gauge\ }}

The time-dependent vector potential generates an electric field%
\begin{equation*}
\mathbf{E}=-\frac{\partial \mathbf{A}}{\partial t},
\end{equation*}%
and thus the Hamiltonian has an electric dipole interaction 
\begin{equation*}
H_{d}=H_{a}+H_{i}^{d},
\end{equation*}%
where%
\begin{equation}
H_{i}^{d}=e\mathbf{x\cdot }\frac{\partial \mathbf{A}}{\partial t}.
\label{10-1}
\end{equation}

The Hamiltonian of second quantization is%
\begin{eqnarray}
\widehat{H_{i}}^{d}\left( t\right) &=&\widehat{P}H_{i}^{d}\widehat{P}%
=G\omega \left( \widehat{a}e^{-i\omega t}-\widehat{a}^{\dag }e^{i\omega
t}\right)  \notag \\
&&\cdot \left( |u_{3}\rangle \left\langle u_{2}\right\vert +|u_{2}\rangle
\left\langle u_{1}\right\vert -c.c\right) ,  \label{11}
\end{eqnarray}%
in which we assume that $\mathbf{e\cdot x}_{ii}=0$ for $i=1,2,3$ due to the
parity symmetry of atom energy levels. Following the same procedure as in
the Coulomb gauge the time-independent Hamiltonian of dipole gauge is 
\begin{equation}
\widehat{H}_{d}=\widehat{H}_{0}+\widehat{H_{i}}^{d},  \label{12}
\end{equation}%
with the dipole gauge interaction given by%
\begin{equation}
\widehat{H_{i}}^{d}=\frac{G\omega }{\sqrt{2}}\left( \widehat{a}-\widehat{a}%
^{\dag }\right) \left( \widehat{s}_{+}-\widehat{s}_{-}\right) ,  \label{13}
\end{equation}%
which is quantitatively different from that in Coulomb gauge Eq. (\ref{10}).
One is proportional to energy difference $\Omega $ of atom while the other
is proportional to the cavity-field frequency $\omega $. This controversy
has lasted for more than half a century

\subsection{\protect\bigskip \emph{The ambiguity of gauge choice}\textbf{\ }}

\bigskip Since the gauge potential $\mathbf{A}\left( t\right) $ is
time-dependent, the total electric field is%
\begin{equation}
\mathbf{E=-\bigtriangledown V-}\frac{\partial \mathbf{A}\left( t\right) }{%
\partial t}.  \label{14}
\end{equation}

The Coulomb gauge Hamiltonian Eq. (\ref{10}) does not include the energy
term of electric field, $\mathbf{-}\frac{\partial \mathbf{A}\left( t\right) 
}{\partial t},$ generated by the time-dependent gauge potential. The failure
comes from the misplacing the canonical momentum $\mathbf{p}$\textbf{\ }by
the kinetic momentum $m\overset{\cdot }{\mathbf{x}}$ in Eq. (\ref{1}). This
misplacement would be no problem for the time independent gauge field, while
the dipole interaction is missing for the time-dependent case. The energy
term $e\mathbf{A\cdot p}$ does not exist in the dipole gauge since the
interaction Eq. (\ref{10-1}) is not derived directly from the standard
minimum coupling Hamiltonian Eq. (\ref{2}). Two gauges are both incomplete
description of the atom-field interaction. This controversy can be resolved
by the time-dependent gauge transformation on the time-dependent Schr\"{o}%
dinger equation.

\section{Time-dependent gauge transformation and the Hamiltonian of unified
gauge}

We apply a time-dependent gauge transformation on the Schr\"{o}dinger
equation with the minimum coupling Hamiltonian Eq. (\ref{2}) such that%
\begin{equation*}
i\frac{\partial }{\partial t}|\psi ^{\prime }\rangle =H^{\prime }|\psi
^{\prime }\rangle ,
\end{equation*}%
where the $U(1)$ gauge transformation operator is given by \cite%
{maggiore2005modern}%
\begin{equation}
R\left( t\right) =e^{-ie\mathbf{x}\cdot \mathbf{A}\left( t\right) },
\label{15}
\end{equation}%
with $|\psi ^{\prime }\rangle =R|\psi \rangle $. The Hamiltonian in the new
gauge is%
\begin{equation*}
H^{\prime }=R\left( t\right) HR^{\dag }\left( t\right) -iR\frac{\partial }{%
\partial t}R^{\dag },
\end{equation*}%
which recovers the dipole gauge Hamiltonian exactly \cite{maggiore2005modern}%
\begin{equation*}
H^{\prime }=H_{a}+e\mathbf{x}\cdot \frac{\partial \mathbf{A}\left( t\right) 
}{\partial t}.
\end{equation*}

If we would directly take the second quantization procedure with identity
density operator $\widehat{I}$, the dipole gauge Hamiltonian should be
obtained. Then the $\mathbf{A\cdot p}$ interaction would be lost. We argue
that the gauge equivalence includes not only the Hamiltonian but also the
state vector. Since the Hamiltonian is in the new gauge with the state
vector $|\psi ^{\prime }\rangle =R|\psi \rangle $, inserting the
completeness operator we have%
\begin{equation*}
|\psi ^{\prime }\rangle =R\sum\limits_{i=1,2,3}|u_{i}\rangle \langle
u_{i}|\psi \rangle =\sum\limits_{i=1,2,3}|u_{i}^{^{\prime }}\rangle \langle
u_{i}|\psi \rangle =\widehat{P}^{\prime }|\psi \rangle .
\end{equation*}

The identity density operator in the new gauge becomes $\widehat{P}^{\prime }
$ and thus the second quantization has to be applied with the transformed
operator $\widehat{P}^{\prime }$. The Hamiltonian of unified gauge is then
derived such as%
\begin{eqnarray}
\widehat{H}_{u} &=&\left( \widehat{P}^{\prime }\right) ^{\dagger
}H^{^{\prime }}\widehat{P}^{\prime }  \notag \\
&=&\widehat{P}\left( H+ex\cdot \frac{\partial A(t)}{\partial t}\right) 
\widehat{P}  \notag \\
&=&\widehat{H}_{a}+\widehat{H}_{i}^{c}(t)+\widehat{H}_{i}^{d}(t),
\label{15-1}
\end{eqnarray}%
in which the Hamiltonian $H$ in the second equality is given in Eq. (\ref{2}%
) for the atom in the cavity field. The Hamiltonian of second quantization
contains two parts of interaction. The first one is exactly that of Coulomb
gauge in Eq. (\ref{3}), while the second part is the interaction Hamiltonian
of dipole gauge given in Eq. (\ref{11}). Following the same procedure of
time-dependent gauge transformation Eqs. (\ref{7-1}), (\ref{7-2}) and (\ref%
{7-3}) of the previous section we obtain the time-independent Hamiltonian of
unified gauge 
\begin{equation}
\widehat{H}_{u}=\widehat{H}_{0}+\widehat{H}_{i}^{c}+\widehat{H}_{i}^{d},
\label{16}
\end{equation}%
in which $\widehat{H}_{0}$, $\widehat{H}_{i}^{c}$, and $\widehat{H}_{i}^{d}$
are given respectively in Eqs. (\ref{9}), (\ref{10}) and (\ref{13}). The
Hamiltonian Eq. (\ref{16}) is truly gauge equivalent to the minimum coupling
one Eq. (\ref{2}).

\bigskip

\section{DM Hamiltonian of three-level atoms, macroscopic quantum states and
QPT}

For $N$ identical three-level atoms, we have the constraint%
\begin{equation*}
\sum_{l=1}^{N}\widehat{P}_{l}=N\widehat{P}.
\end{equation*}

\begin{figure}[tbph]
\includegraphics[width=3in]{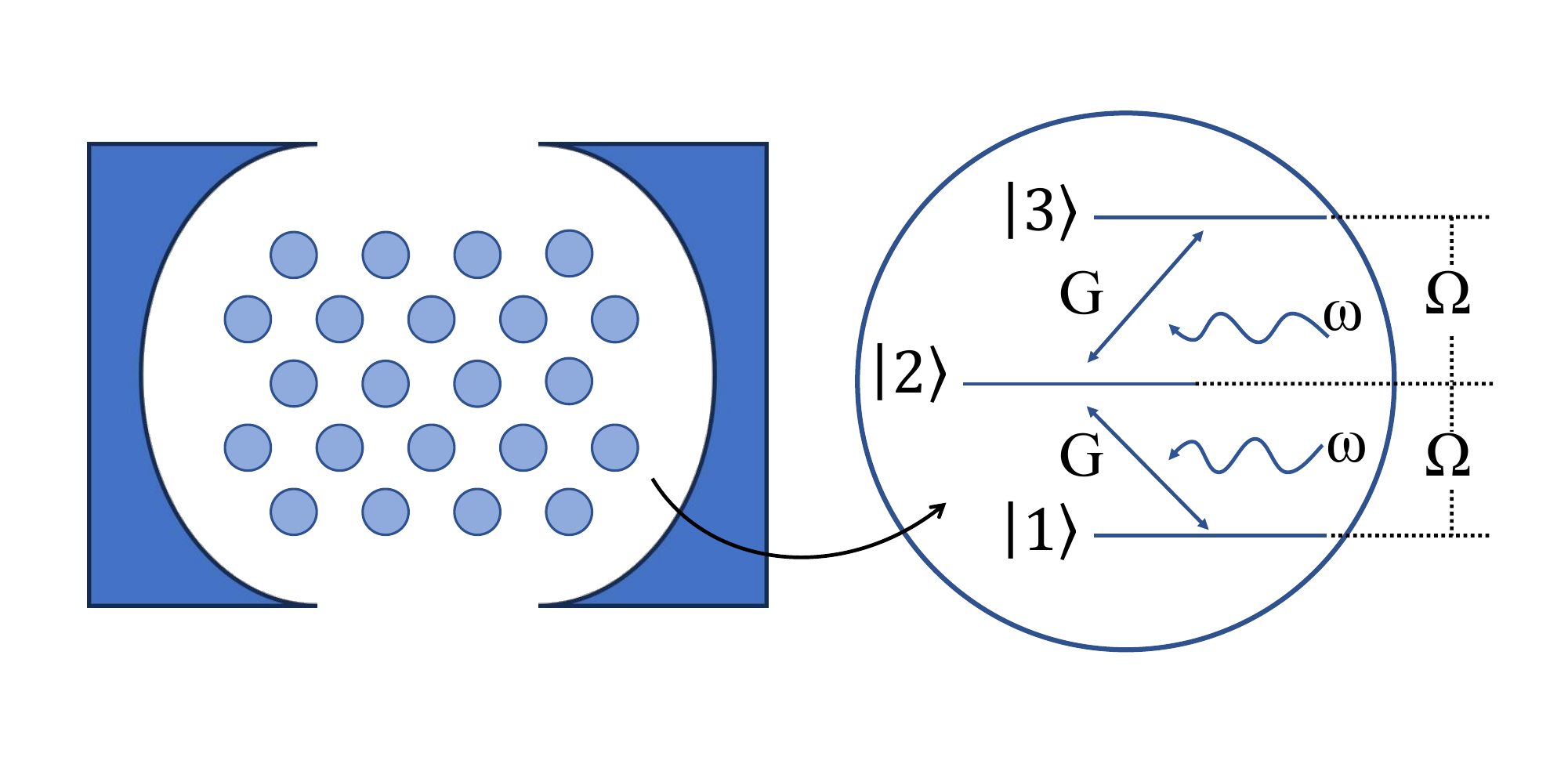}
\caption{Diagram of the extended DM with N three-level atoms in an optical
cavity. Atomic states$\left\vert 1\right\rangle $(bottom), $\left\vert
2\right\rangle $(middle), and $\left\vert 3\right\rangle $(top) are shown by
black horizontal bars. Curved arrows denote photons of frequency $\protect%
\omega $, while solid arrows indicate light-matter interactions between
atomic and cavity modes, with dimensionless coupling constant G between the
light field and atoms}
\label{fig:1}
\end{figure}

Our theoretical model, schematically depicted in Fig. \ref{fig:1}, comprises
N $\Xi $-type three-level atoms coupled to an optical cavity.

The spin-operator Hamiltonian Eq. (\ref{10}) with spin value $1$ is extended
to high spin value $s$. It is crucial to establish the relation between atom
number $N$ and spin value $s.$The number of eigenstates for spin-$s$ is $%
2s+1.$For $N$ identical atoms distributed in three levels there exist $2N+1$
eigenstates. Thus, we have $s=N$ different from the two-level case, where $%
s=N/2$. The DM Hamiltonian for three-level atoms is simply replaced by spin-$%
s$ operators

\bigskip 
\begin{eqnarray}
\widehat{H}_{u} &=&\omega \widehat{a}^{\dag }\widehat{a}+\Omega \widehat{S}%
_{z}+\frac{G\Omega }{\sqrt{2N}}\left( \widehat{a}+\widehat{a}^{\dag }\right)
\left( \widehat{S}_{+}+\widehat{S}_{-}\right)  \notag \\
&&+\frac{G\omega }{\sqrt{2N}}\left( \widehat{a}-\widehat{a}^{\dag }\right)
\left( \widehat{S}_{+}-\widehat{S}_{-}\right) ,  \label{17}
\end{eqnarray}%
where $\widehat{S}_{z}$ and $\widehat{S}_{\pm }$ now are collective, atomic
spin operators.

\subsection{SCS variational method, macroscopic quantum states and QPT}

We are going to demonstrate the macroscopic quantum states and QPT in terms
of SCS variational method \cite%
{PhysRevA.90.023622,PhysRevA.93.033630,Lian_2012,PhysRevA.110.063320}, which
is valid for arbitrary number of atoms $N$. To this end we begin with the
average of Hamiltonian in the optical coherent state $\left\vert \alpha
\right\rangle $, which is assumed as the boson coherent state of cavity mode
such that $a\left\vert \alpha \right\rangle =\alpha \left\vert \alpha
\right\rangle $, where the complex eigenvalue is parametrized as 
\begin{equation}
\alpha =\gamma e^{i\varphi }.  \label{17-2}
\end{equation}%
After the average in the boson coherent state we obtain an effective
Hamiltonian of the pseudo-spin operators 
\begin{equation}
\widehat{H}_{eff}\left( \alpha \right) =\left\langle \alpha \right\vert 
\widehat{H}\left\vert \alpha \right\rangle .  \label{17-3}
\end{equation}

The effective spin Hamiltonian can be diagonalized in the SCSs of north- and
south-pole gauges \cite%
{PhysRevA.90.023622,PhysRevA.93.033630,Lian_2012,PhysRevA.110.063320}
defined by%
\begin{equation}
\widehat{R}\left\vert \pm u\right\rangle =\left\vert s,\pm s\right\rangle
\label{17-1}
\end{equation}%
\begin{equation}
\widehat{R}=e^{-\frac{\theta }{2}\left( \widehat{S}_{+}e^{-i\varphi }-%
\widehat{S}_{-}e^{i\varphi }\right) },  \label{18}
\end{equation}%
where $\theta $ is a pending parameter to be determined. Through a unitary
transformation via the unitary operator $\widehat{R}$, the spin operators $%
\widehat{S}_{z},\widehat{S}_{+},\widehat{S}_{-}$ transform as follows: 
\begin{eqnarray*}
\widehat{R}\widehat{S}_{z}\widehat{R}^{\dag } &=&\widehat{S}_{z}\cos \theta +%
\frac{1}{2}\left( e^{-i\varphi }\widehat{S}_{+}\sin \theta +e^{i\varphi }%
\widehat{S}_{-}\sin \theta \right) \\
\widehat{R}\widehat{S}_{+}\widehat{R}^{\dag } &=&\widehat{S}_{+}\cos ^{2}%
\frac{\theta }{2}-e^{i\varphi }\widehat{S}_{z}\sin \theta -\widehat{S}%
_{-}e^{2i\varphi }\sin ^{2}\frac{\theta }{2} \\
\widehat{R}\widehat{S}_{-}\widehat{R}^{\dag } &=&\widehat{S}_{-}\cos ^{2}%
\frac{\theta }{2}-e^{-i\varphi }\widehat{S}_{z}\sin \theta -\widehat{S}%
_{+}e^{-2i\varphi }\sin ^{2}\frac{\theta }{2}.
\end{eqnarray*}

We in the followings are going to find the energy spectrum of the SCS by
means of the variational method with respect to the variation parameter $%
\gamma $ respectively for three-type of gauge as a comparison.\bigskip

\subsection{Coulomb gauge}

The effective spin Hamiltonian of Coulomb gauge is found after the unitary
transformation as

\begin{equation}
\widehat{R}\widehat{H}_{eff}^{c}\widehat{R}^{\dag }=\omega \gamma ^{2}+A\widehat{S}_{z}+B\widehat{S}_{+}+%
\widehat{C}\widehat{S}_{-},  \label{18-1}
\end{equation}%
in which%
\begin{eqnarray*}
A &=&\Omega \cos \theta -\frac{2\sqrt{2}}{\sqrt{N}}G\Omega \gamma \cos
^{2}\varphi \sin \theta , \\
B &=&\frac{\Omega e^{-i\varphi }}{2}\sin \theta +\frac{\sqrt{2}G\Omega }{%
\sqrt{N}}\gamma \cos \varphi (\cos ^{2}\frac{\theta }{2}-e^{-2i\varphi }\sin
^{2}\frac{\theta }{2}), \\
C &=&\frac{\Omega e^{i\varphi }}{2}\sin \theta +\frac{\sqrt{2}G\Omega }{%
\sqrt{N}}\gamma \cos \varphi (\cos ^{2}\frac{\theta }{2}-e^{2i\varphi }\sin
^{2}\frac{\theta }{2}).
\end{eqnarray*}

The effective spin Hamiltonian Eq. (\ref{18-1}) can be diagonalized with the
SCS Eq. (\ref{17-1}) under the condition $B=0$ and $C=0$, which give rise to 
$\varphi =2n\pi $. And the pending parameter is found as%
\begin{equation*}
\cos \theta =\pm \frac{1}{\sqrt{1+2^{3}\frac{G^{2}}{N}\gamma ^{2}}}.
\end{equation*}

We obtain the semiclassical energy functions in the SCS corresponding
respectively to the ground and highest excited-states 
\begin{equation*}
E_{\pm }^{c}\left( \gamma \right) =\langle \pm u|\widehat{H}_{eff}^{c}|\pm
u\rangle =\omega \gamma ^{2}\pm N\Omega \sqrt{1+\frac{2^{3}G^{2}\gamma ^{2}}{%
N}}.
\end{equation*}

The average energy function of per atom is 
\begin{equation}
\varepsilon _{\pm }\left( \gamma \right) =\frac{E_{\pm }\left( \gamma
\right) }{N}=\frac{\omega \gamma ^{2}}{N}\pm \Omega \sqrt{1+\frac{%
2^{3}\gamma ^{2}G^{2}}{N}}.  \label{25}
\end{equation}

According to the variational method the energies of macroscopic quantum
states are determined from the extremum condition of the energy function%
\begin{equation}
\frac{\partial \varepsilon _{\pm }}{\partial \gamma }=\frac{2\gamma }{N}%
\left( \omega \pm \frac{2^{2}G^{2}\Omega }{\sqrt{1+\frac{2^{3}\gamma
^{2}G^{2}}{N}}}\right) =0.  \label{28}
\end{equation}

The extremum equation Eq. (\ref{28}) always possesses a zero photon-number
solution 
\begin{equation*}
\gamma =0,
\end{equation*}%
which is called the normal phase (NP) if the second-order derivative is
positive 
\begin{equation*}
\frac{\partial ^{2}\varepsilon _{-}}{\partial \gamma ^{2}}\left( \gamma
=0\right) =\frac{2}{N}\left( \omega -4G^{2}\Omega \right) \geq 0.
\end{equation*}

The phase boundary of NP is determined from the equal sign 
\begin{equation}
G_{c}=\frac{1}{2}\sqrt{\eta },  \label{30}
\end{equation}%
in which 
\begin{equation}
\eta =\frac{\omega }{\Omega }  \label{30-1}
\end{equation}%
denotes the ration of field and atom frequencies. The NP with 
\begin{equation*}
\varepsilon _{-}=-\Omega
\end{equation*}%
exists in the region%
\begin{equation*}
G\leq G_{c}.
\end{equation*}

The second-order derivative of the energy function for the spin-up ($%
\varepsilon _{+}$) state is always positive%
\begin{equation*}
\frac{\partial ^{2}\varepsilon _{+}}{\partial \gamma ^{2}}\left( \gamma
=0\right) >0.
\end{equation*}

Thus the zero photon state $\varepsilon _{+}=\Omega $ is stable in the
entire region of coupling value.

The stable macroscopic state of nonzero photon number for energy function $%
\varepsilon _{-}$ is determined from the extremum condition Eq. (\ref{28}).
The nonzero solution is found as%
\begin{equation}
\gamma _{c}=\frac{\sqrt{N}\sqrt{(4G^{2}\frac{1}{\eta })^{2}-1}}{2\sqrt{2}G},
\label{32}
\end{equation}%
which is called the superradiant phase (SP) if the second-order derivative
of the energy function is positive, 
\begin{equation*}
\frac{\partial ^{2}\varepsilon _{-}}{\partial \gamma ^{2}}\left( \gamma
_{c}\right) \geq 0.
\end{equation*}

The equal sign gives rise to exactly the same phase boundary $G_{c}$ in Eq. (%
\ref{30}). The SP exists in the region $G>G_{c}$. The extremum equation Eq. (%
\ref{28}) does not have a nonzero photon solution for the energy function $%
\varepsilon _{+}\left( \gamma \right) $.

\begin{figure}[tbp]
\includegraphics[width=3.5in]{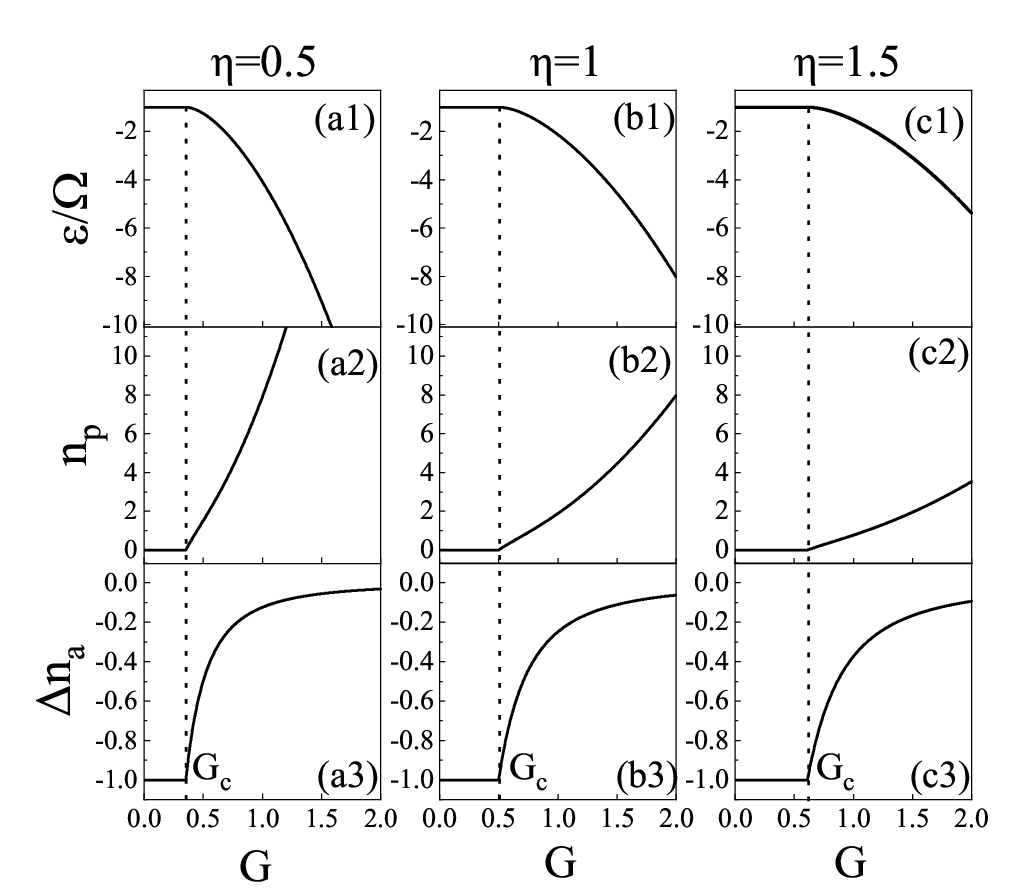}
\caption{Energy spectrum $\protect\varepsilon _{-}$ (1), photon number $n_{p%
\text{ }}$(2), and atomic population $n_{a}$ in the Coulomb gauge
respectively for red detuning $\protect\eta =0.5$ (a), resonance $\protect%
\eta =1$ (b), blue detuning $\protect\eta =1.5$ (c). The phase transition
point $G_{c}$ shifts to the higher value with the increase of detuning. }
\label{fig:2}
\end{figure}

The average photon number per atom, energy and atomic population imbalance
between the two atom levels $\pm \Omega $ in SP are given respectively by%
\begin{equation*}
n_{p}=\frac{\gamma _{c}^{2}}{N}=\frac{(2^{2}G^{2}\frac{1}{\eta })^{2}-1}{%
2^{3}G^{2}},
\end{equation*}%
\begin{equation*}
\varepsilon _{-}=\omega n_{p}-\Omega \sqrt{1+2^{3}n_{p}G^{2}},
\end{equation*}%
and%
\begin{equation*}
\Delta n_{a}=\frac{\left\langle -u\right\vert \widehat{S}_{z}\left\vert
-u\right\rangle }{N}=-\frac{1}{\sqrt{1+2^{3}n_{p}G^{2}}},
\end{equation*}%
which are depicted in fig. \ref{fig:2} with respect to the dimensionless
coupling constant $G$.

\subsection{Dipole gauge}

The effective spin Hamiltonian of dipole gauge $\widehat{H}_{eff}^{d}$ has
the same form of Eq. (\ref{18-1}) after unitary transformation with%
\begin{equation*}
A=\Omega \cos \theta +\frac{2\sqrt{2}G}{\sqrt{N}}\gamma \omega \sin
^{2}\varphi \sin \theta ,
\end{equation*}%
\begin{eqnarray*}
B &=&\frac{\Omega e^{-i\varphi }}{2}\sin \theta +i\frac{\sqrt{2}G\omega }{%
\sqrt{N}}\gamma \sin \varphi \left( \cos ^{2}\frac{\theta }{2}+e^{-2i\varphi
}\sin ^{2}\frac{\theta }{2}\right) , \\
C &=&\frac{\Omega e^{i\varphi }}{2}\sin \theta -i\frac{\sqrt{2}G\omega }{%
\sqrt{N}}\gamma \sin \varphi (e^{2i\varphi }\sin ^{2}\frac{\theta }{2}+\cos
^{2}\frac{\theta }{2}).
\end{eqnarray*}

The pending parameter $\theta $ is determined under the condition $B=C=0$,

\begin{equation}
\cos \theta =\pm \frac{1}{\sqrt{1+2^{3}G^{2}\frac{\gamma ^{2}}{N}\eta ^{2}}},
\label{40}
\end{equation}%
with the field phase given by $\varphi =\frac{\pi }{2}+2n\pi $. Finally the
average energy functions of per atom are found in the SCS $|\pm u>$ as%
\begin{equation*}
\varepsilon _{\pm }=\frac{\omega \gamma ^{2}}{N}\pm \Omega \sqrt{1+\frac{%
2^{3}\gamma ^{2}G^{2}\eta ^{2}}{N}}.
\end{equation*}

The extremum equation

\begin{equation}
\frac{\partial \varepsilon _{\pm }}{\partial \gamma }=\frac{2\gamma \omega }{%
N}\left( 1\pm \frac{2^{2}G^{2}\eta }{\sqrt{1+\frac{2^{3}\gamma ^{2}G^{2}\eta
^{2}}{N}}}\right) =0  \label{46}
\end{equation}%
always possesses a zero photon-number solution $\gamma =0$.

The second-order derivative of the energy function for the spin-up state $%
\left( \varepsilon _{+}\right) $ is always positive%
\begin{equation*}
\frac{\partial ^{2}\varepsilon _{+}}{\partial \gamma ^{2}}\left( \gamma
=0\right) =\frac{2\omega }{N}\left( 1+2^{2}G^{2}\eta \right) >0,
\end{equation*}%
which means that the macroscopic quantum state $\varepsilon _{+}$ is stable
same as in the Coulomb gauge.

The nonzero photon solution for the spin-down state is obtained from the
extremum equation

\begin{equation}
\frac{\partial \varepsilon _{-}}{\partial \gamma }=\frac{2\gamma \omega }{N}%
\left( 1-\frac{2^{2}G^{2}\eta }{\sqrt{1+\frac{2^{3}\gamma ^{2}G^{2}\eta ^{2}%
}{N}}}\right) =0,  \label{48}
\end{equation}%
which gives rise to the zero photon solution $\gamma =0$ and the nonzero
photon solution 
\begin{equation}
\gamma _{c}=\frac{\sqrt{N}}{2\sqrt{2}}\sqrt{2^{4}G^{2}-\left( \frac{1}{G\eta 
}\right) ^{2}}.  \label{49}
\end{equation}

The stable zero photon solution satisfies the condition%
\begin{equation*}
\frac{\partial ^{2}\varepsilon _{-}}{\partial \gamma ^{2}}\left( \gamma
=0\right) =\frac{2\omega }{N}\left( 1-2^{2}G^{2}\eta \right) \geq 0.
\end{equation*}

The NP exists in the region%
\begin{equation*}
G\leq G_{c},
\end{equation*}%
with the phase boundary given by%
\begin{equation}
G_{c}=\frac{1}{2}\sqrt{\frac{1}{\eta }},  \label{49-1}
\end{equation}%
which is different from that of Coulomb gauge Eq.(\ref{30}). The stable
nonzero-photon state, namely the SP, is verified by the positive
second-order derivative 
\begin{equation*}
\frac{\partial ^{2}\varepsilon _{-}}{\partial \gamma ^{2}}\left( \gamma
_{c}\right) \geq 0,
\end{equation*}%
in which the equality results also in the phase boundary Eq.(\ref{49-1})
between NP and the SP. The average photon number of per atom, energy and
atomic population imbalance are given respectively by%
\begin{equation}
n_{p}=\frac{\gamma _{c}^{2}}{N}=\frac{1}{2^{3}}\left[ 2^{4}G^{2}-\left( 
\frac{1}{G\eta }\right) ^{2}\right] ,  \label{50}
\end{equation}%
\begin{equation}
\varepsilon _{-}=\omega n_{p}\pm \Omega \sqrt{1+2^{3}n_{p}G^{2}\eta ^{2}},
\label{50-1}
\end{equation}%
and%
\begin{equation*}
\Delta n_{a}=-\frac{1}{\sqrt{1+2^{3}n_{p}G^{2}\eta ^{2}}},
\end{equation*}%
which reduces to the well known result in the NP that $\Delta n_{a}=-1$.

\begin{figure}[tbp]
\includegraphics[width=3.5in]{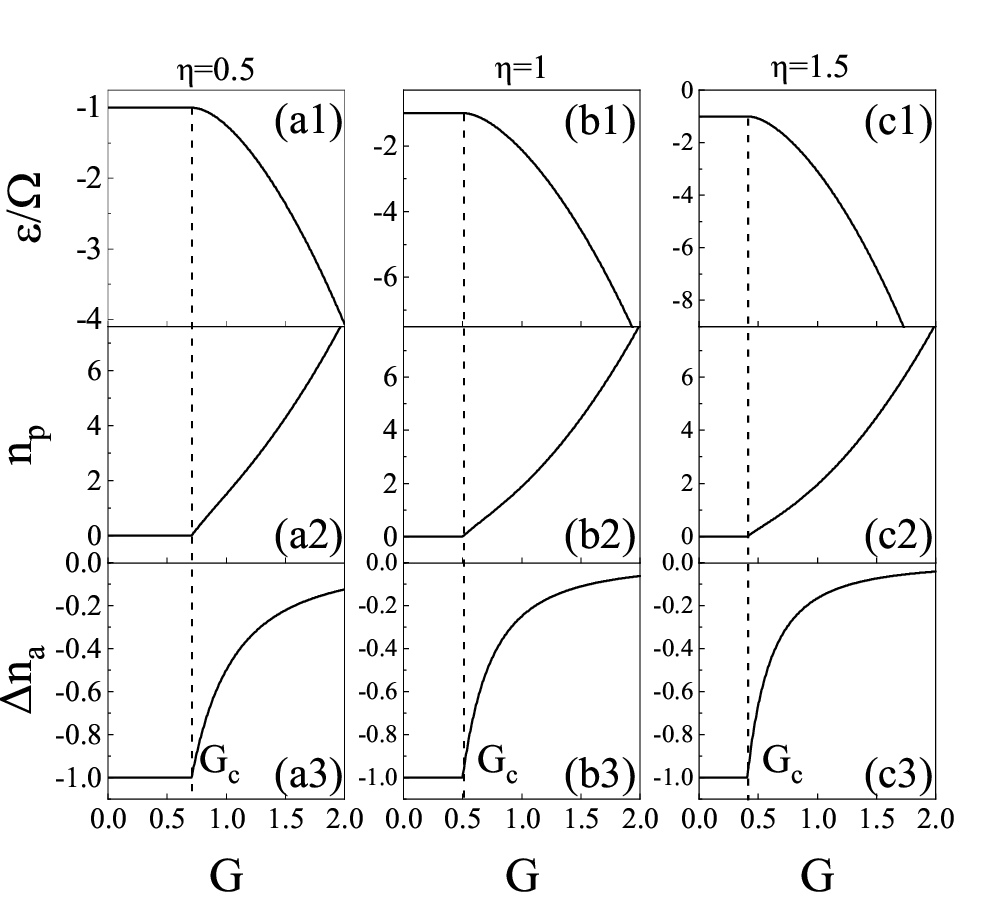}
\caption{The energy spectrum $\protect\varepsilon _{-}$ (1), photon number $%
n_{p}$ (2) and atomic population $\Delta n_{a}$ (3) in the dipole gauge
respectively for red-detuning (a), resonance (b), and blue-detuning (c). The
phase transition point $G_{c}$ shifts to the lower value with the increase
of detuning.}
\label{fig:3}
\end{figure}

\begin{figure}[th]
\centering
\includegraphics[width=3in]{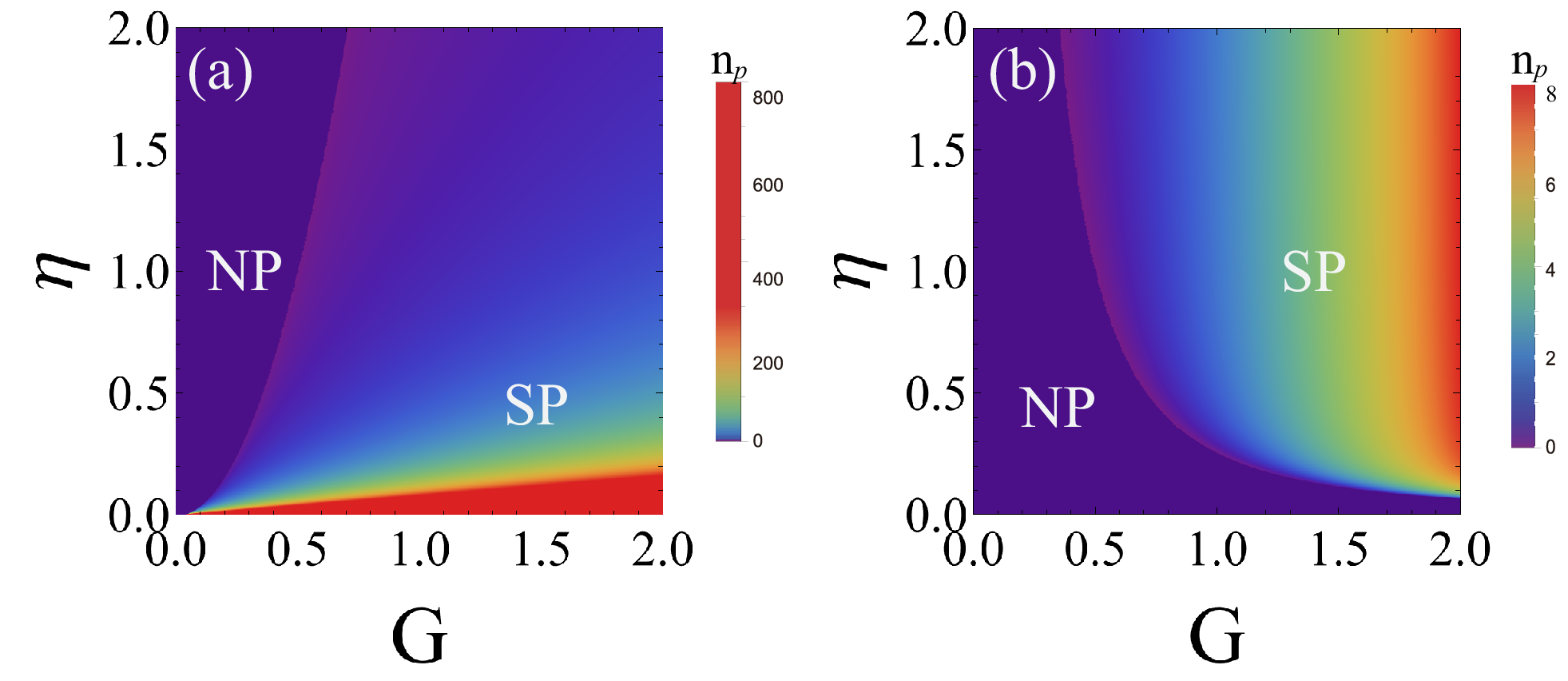}
\caption{Phase diagrams in $G-\protect\eta $ space for Coulomb gauge (a) and
dipole gauge (b). The photon number $n_{p}$ in SP is indicated by color
scale.}
\label{fig:4}
\end{figure}

Corresponding quantities ($\varepsilon _{-}$, $n_{p}$, $\Delta n_{a}$) in
the dipole gauge are shown in Fig. \ref{fig:3}. The phase transition point $%
G_{c}$ moves to the opposite direction with increase of the atom-field
frequency detuning respectively in the Coulomb and dipole gauges, while has
equal value $\frac{1}{2}$ in the resonance $\eta =1$. Fig. \ref{fig:4} is
the phase diagram in $G-$ $\eta $ space for both gauges. Phase boundary 
separates the NP (dark purple region) and SP with the photon number $n_{p}$
in color scale. Critical observation is that the SP region decreases
(increases) with the increasing detuning $\eta $ in Coulomb (dipole) gauge
consistent with Figs. \ref{fig:2} and \ref{fig:3}.

\subsection{Unified gauge}

For the effective spin Hamiltonian Eq. (\ref{17-3}) in the unified gauge the
unitary transformation operator Eq. (\ref{18}) has to be modified as%
\begin{equation*}
\widehat{R}=e^{-\frac{\theta }{2}\left( \widehat{S}_{+}e^{-i\phi }-\widehat{S%
}_{-}e^{i\phi }\right) },
\end{equation*}%
in which $\theta $, $\phi $ are two pending parameters to be determined.
After the unitary transformation the effective spin Hamiltonian has the same
form of Eq. (\ref{18-1}) with

\begin{eqnarray*}
B &=&G\sqrt{\frac{2}{N}}\gamma \lbrack \Omega \cos \varphi (\cos ^{2}\frac{%
\theta }{2}-e^{-2i\phi }\sin ^{2}\frac{\theta }{2}) \\
&&+i\omega \sin \varphi (\cos ^{2}\frac{\theta }{2}+e^{-2i\phi }\sin ^{2}%
\frac{\theta }{2})]+\frac{1}{2}\Omega \sin \theta e^{-i\phi },
\end{eqnarray*}%
\begin{eqnarray*}
C &=&G\sqrt{\frac{2}{N}}\gamma \lbrack \Omega \cos \varphi (\cos ^{2}\frac{%
\theta }{2}-e^{i2\phi }\sin ^{2}\frac{\theta }{2}) \\
&&-i\omega \sin \varphi (e^{i2\phi }\sin ^{2}\frac{\theta }{2}+\cos ^{2}%
\frac{\theta }{2})]+\frac{1}{2}\Omega \sin \theta e^{i\phi },
\end{eqnarray*}%
and%
\begin{equation*}
A=\Omega \cos \theta +\frac{2\sqrt{2}\gamma }{\sqrt{N}}G\sin \theta (\omega
\sin \varphi \sin \phi -\Omega \cos \varphi \cos \phi ).
\end{equation*}%
The pending parameters $\phi $ and $\theta $ are determined by the condition 
$B=C=0$ that 
\begin{equation*}
\cos \phi =\frac{\cos \varphi }{\sqrt{\Phi \left( \eta ,\varphi \right) }},
\end{equation*}%
in which 
\begin{equation*}
\Phi \left( \eta ,\varphi \right) =\cos ^{2}\varphi +\eta ^{2}\sin
^{2}\varphi ,
\end{equation*}%
is a function of cavity-field phase $\varphi $ in Eq.(\ref{17-2}). We have%
\begin{equation*}
\cos \theta =\pm \frac{1}{\sqrt{1+2^{3}G^{2}\frac{\gamma ^{2}}{N}\Phi \left(
\eta ,\varphi \right) }},
\end{equation*}%
and 
\begin{equation}
A=\Omega \sqrt{1+2^{3}G^{2}\frac{\gamma ^{2}}{N}\Phi \left( \eta ,\varphi
\right) }.  \label{50-2}
\end{equation}

The semiclassical energy function of per atom in the SCS $|\pm u>$ depends
on the field phase $\varphi $%
\begin{equation}
\varepsilon _{\pm }\left( \gamma ,\varphi \right) =\omega \frac{\gamma ^{2}}{%
N}\pm A,  \label{50-3}
\end{equation}%
which reduces to the energy functions of Coulomb and dipole gauges
respectively in $\varphi =0,\frac{\pi }{2}$. The extremum equation 
\begin{equation}
\frac{\partial \varepsilon _{\pm }}{\partial \gamma }=0  \label{51}
\end{equation}%
possesses the zero photon solution $\gamma =0$, which is stable for the spin
down state $\varepsilon _{-}$ under the condition 
\begin{equation*}
\frac{\partial ^{2}\varepsilon _{-}}{\partial \gamma ^{2}}(\gamma =0)\geq 0.
\end{equation*}

The boundary condition is found from the equality 
\begin{equation}
G_{c}=\frac{1}{2}\sqrt{\frac{\eta }{\Phi \left( \eta ,\varphi \right) }},
\label{52}
\end{equation}%
which is the same value $G_{c}=\frac{1}{2}$ in the resonance $\eta =1$
independent of the gauge choice. The spin down state with energy $%
\varepsilon _{-}$ is called the NP in the region $G<$ $G_{c}$. The spin up
state of $\varepsilon _{+}$ with population inversion is stable since the
second derivative of energy function, $\frac{\partial ^{2}\varepsilon _{+}}{%
\partial \gamma ^{2}}(\gamma =0)>0$, is always greater than $0$.

The nonzero photon solution of the extremum equation Eq. (\ref{50}) for spin
down state given by

\begin{equation}
\gamma _{c}=\sqrt{N}\sqrt{\frac{2G^{2}\Phi \left( \eta ,\varphi \right) }{%
\eta ^{2}}-\frac{1}{2^{3}G^{2}\Phi \left( \eta ,\varphi \right) }}.
\label{53}
\end{equation}

\begin{figure}[]
\centering
\includegraphics[width=3in]{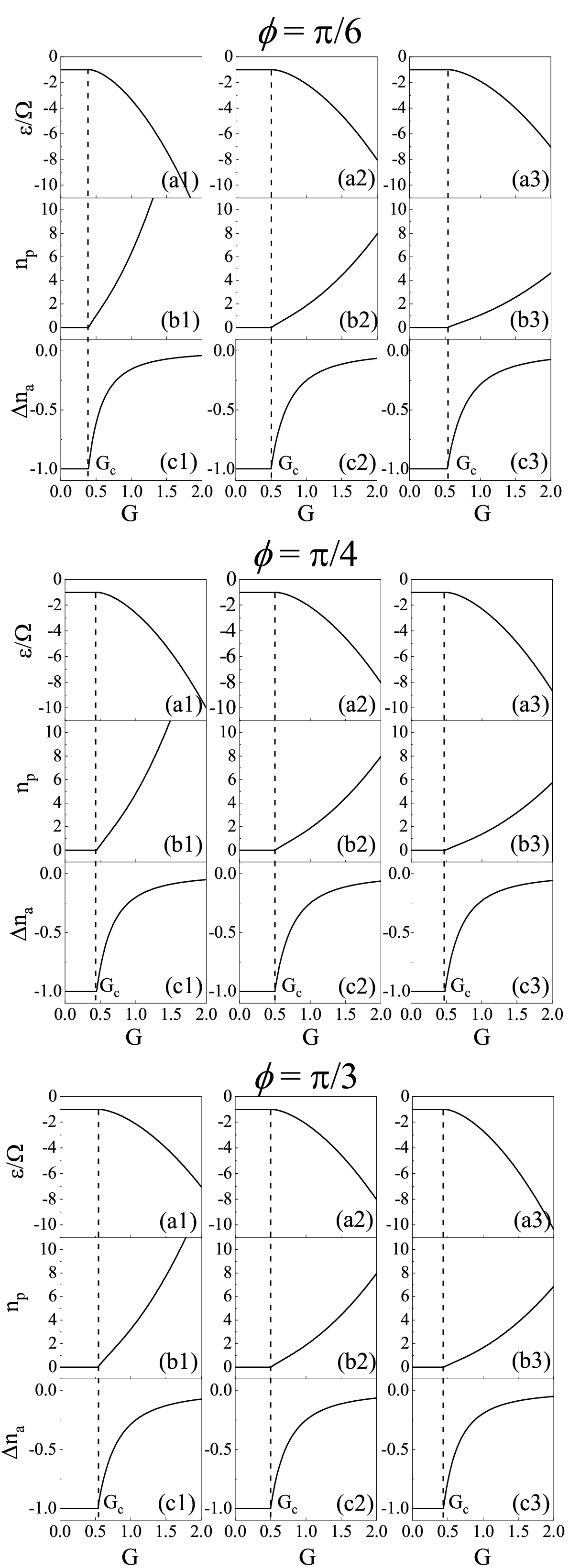}
\caption{The energy spectrum $\protect\varepsilon _{-}$ (a), photon number $%
n_{p}$ (b) and atomic population $\Delta n_{a}$ (c) at fixed field phases $%
\protect\phi $ =$\protect\pi /6,\protect\pi /4,\protect\pi /3$ for red
detuning $\protect\eta =0.5$ (1), resonance $\protect\eta =1$ (2), blue
detuning $\protect\eta =1.5$ (3). }
\label{fig:5}
\end{figure}

The stable ground state, namely the SP, is verified by the positive
second-order derivative 
\begin{equation*}
\frac{\partial ^{2}\varepsilon _{-}}{\partial \gamma ^{2}}\left( \gamma
_{c}\right) \geq 0,
\end{equation*}%
in which the equality equation results also in the phase boundary Eq. (\ref%
{52}) between NP and the SP. The average photon number of per atom, energy
and atomic population imbalance for the SP are respectively given by%
\begin{equation*}
n_{p}=\frac{\gamma _{c}^{2}}{N}
\end{equation*}%
\begin{equation}
\varepsilon _{-}\left( n_{p}\right) =\omega n_{p}-\Omega \sqrt{%
1+2^{3}G^{2}n_{p}\Phi \left( \eta ,\varphi \right) }  \label{55}
\end{equation}%
and%
\begin{equation}
\Delta n_{a}=-\frac{1}{\sqrt{1+2^{3}G^{2}n_{p}\Phi \left( \eta ,\varphi
\right) }},  \label{56}
\end{equation}%
which reduce to those in Coulomb gauge for $\varphi =0$ and dipole gauge for 
$\varphi =\pi /2$. These quantities are plotted in Fig. \ref{fig:5}
respectively for field phases $\varphi =\pi /6,\pi /4,\pi /3$ indicating the
phase-sensitive behavior. They are the same at resonance independent of the
gauge and the phase angle $\varphi $. The increase of photon number with
coupling $G$ slows slightly when phase angle $\varphi $ increases in the red
detuning seen from Fig. \ref{fig:5} (b1). While the situation is just
opposite in the blue detuning [Fig. \ref{fig:5} (b3)].

\begin{figure*}[th]
\centering
\includegraphics[width=6in]{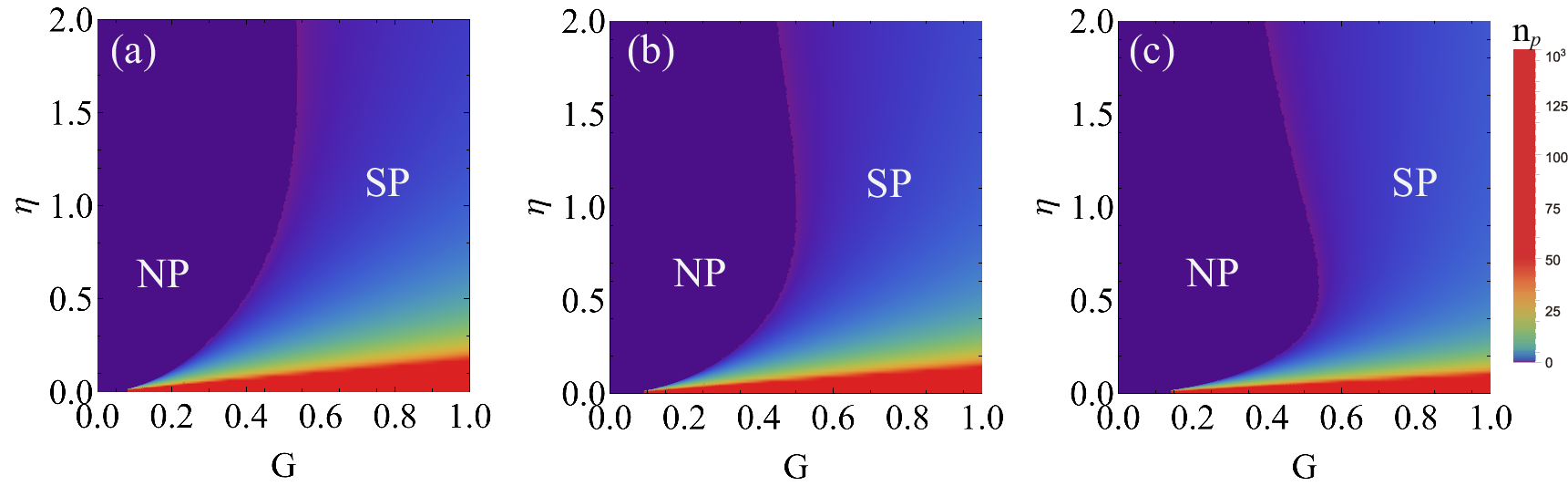}
\caption{Phase diagrams in $G-\protect\eta $ space for cavity-field phases $%
\protect\varphi =\protect\pi /6$ (a), $\protect\pi /4$ (b) , $\protect\pi /3$
(c).}
\label{fig:6}
\end{figure*}

Fig. \ref{fig:6} presents $G-\eta $ phase diagrams in the unified gauge for
different optical phases $\varphi =\pi /6,\pi /4,\pi /3$ indicating the
continuos variation of the phase boundary and photon number in SP with
respect to $G$ and $\eta $.

The geometric phase induced by the light field can be found by solving the
Schr\H{o}dinger equation with the time-dependent Hamiltonian Eq. (\ref{15-1}%
).

Using the unitary transformation Eq. (\ref{15}) the non-adiabatic Berry
phase in the ground state $|\psi \rangle =|\alpha \rangle |-u\rangle $ is
found as \cite{Liang2023,PhysRevA.74.054101} 
\begin{equation}
\Gamma =i\int_{0}^{T}\left\langle \psi \right\vert \widehat{U}\frac{\partial 
}{\partial t}\widehat{U}^{\dag }\left\vert \psi \right\rangle dt=2\pi
\left\langle \alpha \right\vert \widehat{a}^{\dag }\widehat{a}\left\vert
\alpha \right\rangle =2\pi \gamma ^{2}.  \label{76}
\end{equation}

The average geometric phase of per atom is zero in NP and

\begin{equation*}
\Upsilon =\frac{\Gamma }{N}=2\pi n_{p},
\end{equation*}%
in the SP. It is well known that the Berry phase can characterize the QPT
and the critical property at the phase transition point $G_{c}$.

\section{Non-Hermitian Hamiltonian, exceptional point and ground state}

The non-Hermitian Hamiltonian has become a hot topic in resent years. The
interaction constant $G$ in the DM is proportional to dipole matrix element $%
\mathbf{x}_{ij}$, which is a complex number with equal order of real and
imaginary parts \cite{peng1998introduction}. The Hermitian interaction is
obtained with the imaginary value of $\mathbf{x}_{ij}$. It is certainly
interesting to see the phenomena of non-Hermitian cavity-atom coupling with
the real part of the matrix element $\mathbf{x}_{ij}$ Eq. (\ref{5-1}). The
non-Hermitian Hamiltonian of DM can be expressed as%
\begin{eqnarray}
\widehat{H}_{u} &=&\omega \widehat{a}^{\dag }\widehat{a}+\Omega \widehat{S}%
_{z}+\frac{iG}{\sqrt{2N}}[\Omega \left( \widehat{a}+\widehat{a}^{\dag
}\right) \left( \widehat{S}_{+}+\widehat{S}_{-}\right)  \notag \\
&&+\omega \left( \widehat{a}-\widehat{a}^{\dag }\right) \left( \widehat{S}%
_{+}-\widehat{S}_{-}\right) ].  \label{78}
\end{eqnarray}

After the average in the boson coherent state we obtain an effective
Hamiltonian of the pseudospin operators in the unified gauge%
\begin{eqnarray}
\widehat{H}_{eff} &=&\omega \gamma ^{2}+\Omega \widehat{S}_{z}+\frac{\sqrt{2}%
G\gamma }{\sqrt{N}}[i\Omega \cos \varphi \left( \widehat{S}_{+}+\widehat{S}%
_{-}\right)  \notag \\
&&-\omega \sin \varphi \left( \widehat{S}_{+}-\widehat{S}_{-}\right) ],
\label{79}
\end{eqnarray}%
which is non-Hermitian%
\begin{eqnarray}
\widehat{H}_{eff}^{\dag } &=&\omega \gamma ^{2}+\Omega \widehat{S}_{z}+\frac{%
\sqrt{2}G\gamma }{\sqrt{N}}[-i\Omega \cos \varphi \left( \widehat{S}_{+}+%
\widehat{S}_{-}\right)  \notag \\
&&+\omega \sin \varphi \left( \widehat{S}_{+}-\widehat{S}_{-}\right) ].
\label{80}
\end{eqnarray}

According to the theory of Mostafazadeh \cite%
{Mostafazadeh2002,Mostafazadeh2002b,Mostafazadeh2007} the non-Hermitian
Hamiltonian with real spectrum can be transformed to a Hermitian one by a\
similarity transformation. This is called the pseudo-symmetry, which in the
present case is%
\begin{equation}
\widehat{R}\widehat{H}_{eff}\widehat{R}^{-1}=\widehat{h}=\widehat{h}^{\dag }.
\label{81}
\end{equation}

The Hermitian transformation operator is seen to be \cite%
{PhysRevA.110.063320}%
\begin{equation}
\widehat{R}=e^{-\frac{\theta }{2}\left( \widehat{S}_{+}e^{i\phi }+\widehat{S}%
_{-}e^{-i\phi }\right) }.  \label{84}
\end{equation}

Using the transformation form%
\begin{eqnarray*}
\widehat{R}\widehat{S}_{z}\widehat{R}^{-1} &=&\widehat{S}_{z}\cosh \theta +%
\frac{1}{2}e^{i\phi }\widehat{S}_{+}\sinh \theta -\frac{1}{2}e^{-i\phi }%
\widehat{S}_{-}\sinh \theta , \\
\widehat{R}\widehat{S}_{+}\widehat{R}^{-1} &=&\widehat{S}_{+}\cosh ^{2}\frac{%
\theta }{2}+e^{-i\phi }\widehat{S}_{z}\sinh \theta -\widehat{S}%
_{-}e^{-2i\phi }\sinh ^{2}\frac{\theta }{2}, \\
\widehat{R}\widehat{S}_{-}\widehat{R}^{-1} &=&\widehat{S}_{-}\cosh ^{2}\frac{%
\theta }{2}-e^{i\phi }\widehat{S}_{z}\sinh \theta -\widehat{S}_{+}e^{2i\phi
}\sinh ^{2}\frac{\theta }{2}.
\end{eqnarray*}%
The effective spin Hamiltonian $\widehat{H}_{eff}$ possesses the same form
Eq. (\ref{18-1}) with%
\begin{equation*}
A=\Omega \cosh \theta +\frac{2\sqrt{2}G\gamma }{\sqrt{N}}(\Omega \cos
\varphi \sin \phi -\omega \sin \varphi \cos \phi )\sinh \theta ,
\end{equation*}%
\begin{eqnarray*}
B &=&\frac{\sqrt{2}G\gamma }{\sqrt{N}}[i\Omega \cos \varphi (\cosh ^{2}\frac{%
\theta }{2}-e^{2i\phi }\sinh ^{2}\frac{\theta }{2}) \\
&&-\omega \sin \varphi (\cosh ^{2}\frac{\theta }{2}+e^{2i\phi }\sinh ^{2}%
\frac{\theta }{2})]+\frac{\Omega }{2}e^{i\phi }\sinh \theta ,
\end{eqnarray*}%
and%
\begin{eqnarray*}
C &=&\frac{\sqrt{2}G\gamma }{\sqrt{N}}[i\Omega \cos \varphi (\cosh ^{2}\frac{%
\theta }{2}-e^{-2i\phi }\sinh ^{2}\frac{\theta }{2}) \\
&&+\omega \sin \varphi (\cosh ^{2}\frac{\theta }{2}+e^{-2i\phi }\sinh ^{2}%
\frac{\theta }{2})]-\frac{\Omega }{2}e^{-i\phi }\sinh \theta .
\end{eqnarray*}

The Hermitian Hamiltonian can be indeed obtained under the condition $B=C=0$%
, from which the pending parameters is determined as%
\begin{equation*}
\cos \phi =\frac{\eta \sin \varphi }{\sqrt{\Phi \left( \eta ,\varphi \right) 
}},
\end{equation*}%
and 
\begin{equation*}
\cosh \theta =\pm \frac{1}{\sqrt{1-2^{3}G^{2}\frac{\gamma ^{2}}{N}\Phi
\left( \eta ,\varphi \right) }}.
\end{equation*}

We obtain%
\begin{equation}
A=\Omega \sqrt{1-2^{3}G^{2}\frac{\gamma ^{2}}{N}\Phi \left( \eta ,\varphi
\right) }.  \label{85}
\end{equation}

The Hamiltonian after transformation \ 
\begin{equation}
\widehat{h}=\omega \gamma ^{2}+A\widehat{S}_{z},  \label{86-1}
\end{equation}%
is Hermitian under condition%
\begin{equation*}
2^{3}G^{2}\frac{\gamma ^{2}}{N}\Phi \left( \eta ,\varphi \right) \leq 1.
\end{equation*}

The Hamiltonian Eq. (\ref{86-1}) has eigenstates $\widehat{h}|\pm s\rangle
=\pm s|\pm s\rangle $ with $s=N$ being the extremum SCSs. The eigenstates of
original non-Hermitian Hamiltonians are seen to be form Eq. (\ref{81})

\begin{eqnarray*}
\widehat{H}_{eff}|\pm u\rangle _{r} &=&E_{\pm }|\pm u\rangle _{r}, \\
\widehat{H}_{eff}^{\dag }|\pm u\rangle _{l} &=&E_{\pm }|\pm u\rangle _{l},
\end{eqnarray*}%
in which 
\begin{eqnarray*}
|\pm u\rangle _{r} &=&\widehat{R}^{-1}|\pm s\rangle , \\
|\pm u\rangle _{l} &=&\widehat{R}|\pm s\rangle ,
\end{eqnarray*}%
are called respectively the ``ket" and ``bra" states related by a metric
operator%
\begin{equation*}
\widehat{\chi }=\widehat{R}^{2}
\end{equation*}%
such that%
\begin{equation*}
|\pm u\rangle _{l}=\widehat{\chi }|\pm u\rangle _{r}.
\end{equation*}

The semiclassical energy functions are obtained as%
\begin{equation}
E_{\pm }\left( \gamma \right) =_{l}\langle \pm u|\widehat{H}_{eff}|\pm
u\rangle _{r}=N\varepsilon _{\pm }\left( \gamma \right) ,  \label{87-1}
\end{equation}%
with 
\begin{equation}
\varepsilon _{\pm }\left( \gamma \right) =\omega \frac{\gamma ^{2}}{N}\pm A.
\label{87-2}
\end{equation}

\subsection{Exceptional point of semiclassical energy function}

An important observation for the non-Hermitian Hamiltonian is that the
energy functions Eq. (\ref{87-1}) are not always real. An exceptional point
(EP)%
\begin{equation}
G_{ep}=\frac{\sqrt{N}}{2\sqrt{2}\gamma }\frac{1}{\sqrt{\Phi \left( \eta
,\varphi \right) }},  \label{88}
\end{equation}%
exists. The energy function is real below $G_{ep}$, while becomes complex
when $G>$ $G_{ep}$.

\begin{figure}[th]
\centering
\begin{subfigure}[c]{0.8\linewidth}
		\centering
	 \begin{overpic}[width=\linewidth]{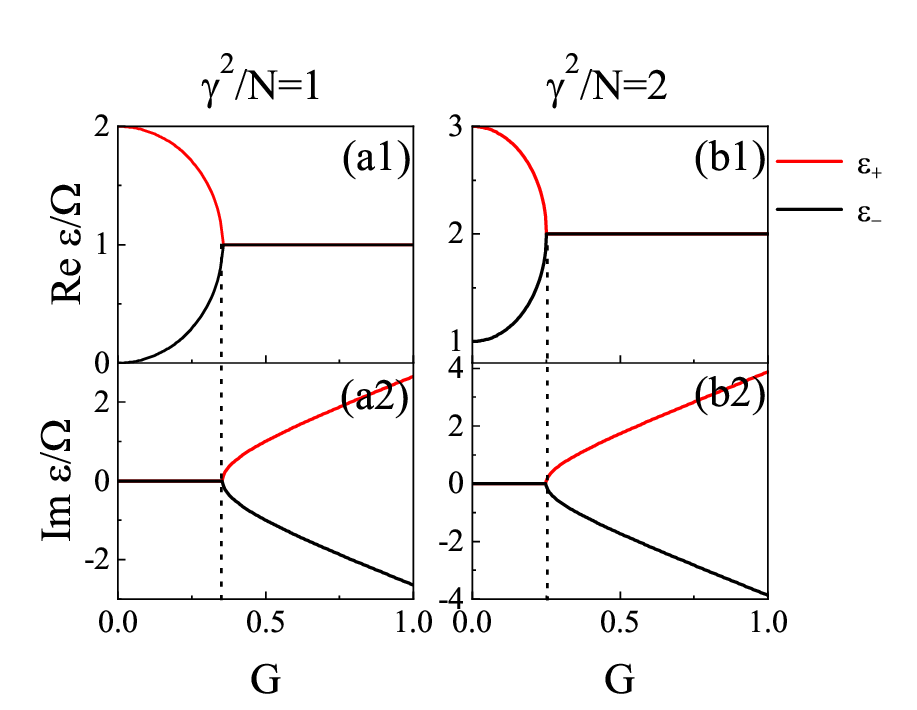}
        \end{overpic}
\end{subfigure}
\caption{ Real (1) and imaginary (2) parts of energy functions $\protect%
\varepsilon _{\pm }\left( \protect\gamma \right) $ as a function of the
dimensionless atom-field coupling strength $G$ for the parameter $\protect%
\gamma ^{2}/N=1$ (a), $2$ (b) with $\protect\eta =1$, $\protect\varphi =%
\protect\pi /3$. }
\label{fig:7}
\end{figure}

The variations of real part (1) and imaginary part (2) of energy functions $%
\varepsilon _{\pm }\left( \gamma \right) $ Eq. (\ref{87-1}) with respect to
the atom-field coupling $G$ are displayed in Fig. \ref{fig:7} under the
resonance $\eta =1$ with given cavity-field parameter $\gamma ^{2}/N=1$ (a), 
$2$ (b). Two branches of the energy function $\varepsilon _{\pm }\left(
\gamma \right) $ are represented respectively by the red and black solid
lines. In the region $G<G_{ep}$, the real part of energy functions have two
branches, while the imaginary part is zero. However, the situation is just
opposite for $G>G_{ep}$.

\begin{figure}[th]
\centering
\begin{subfigure}[c]{0.8\linewidth}
		\centering
	 \begin{overpic}[width=\linewidth]{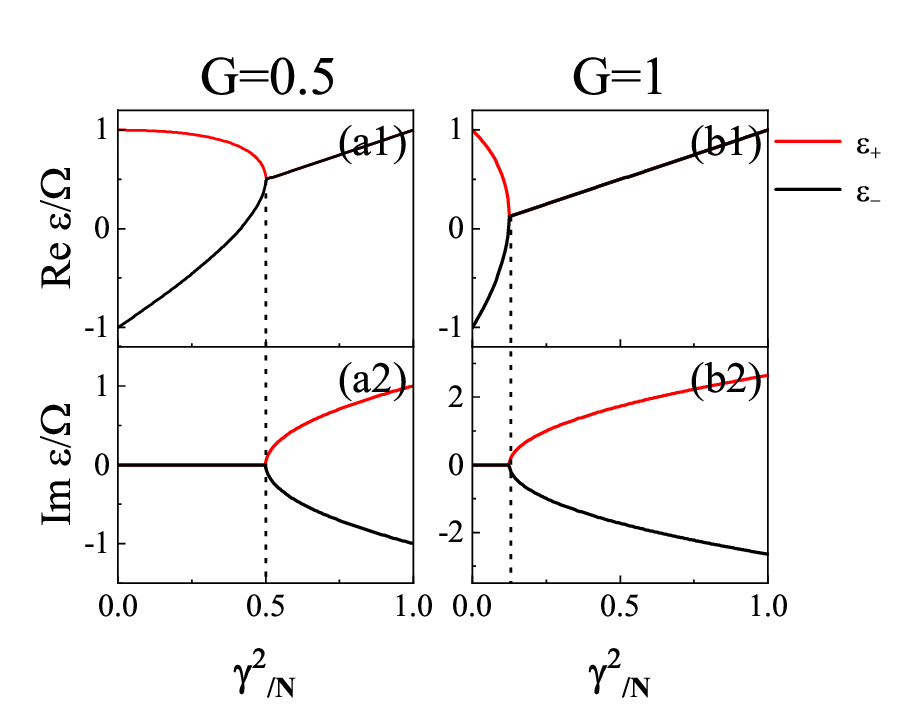}
        \end{overpic}
\end{subfigure}
\caption{The variations of real (1) and imaginary (2) parts of energy
functions $\protect\varepsilon _{\pm }\left( \protect\gamma \right) $ with
respect to the cavity-field parameter $\protect\gamma ^{2}/N$ for the
atom-field coupling strength $G=0.5$ (a), $1$ (b) with $\protect\eta =1$, $%
\protect\varphi =\protect\pi /3$. }
\label{fig:8}
\end{figure}

Fig. \ref{fig:8} shows the variations of energy functions $\varepsilon _{\pm
}\left( \gamma \right) $ with respect to the cavity-field parameter $\gamma
^{2}/N$ under the resonance ($\eta =1$).

\subsection{Variational ground state}

We now investigate the variational ground state with the energy function
given by Eq. (\ref{87-2}) 
\begin{equation*}
\varepsilon _{\pm }\left( \gamma \right) =\frac{\omega \gamma ^{2}}{N}\pm
\Omega \sqrt{1-\frac{2^{3}G^{2}\gamma ^{2}}{N}\Phi \left( \eta ,\varphi
\right) }.
\end{equation*}

By the variation with respect to parameter $\gamma $, we find that the spin
down state $\varepsilon _{-}\left( \gamma =0\right) $ of zero photon is
stable in the whole region of $G$ different from the Hermitian case where
the NP exists only in lower value region. While the stable spin-up state $%
\varepsilon _{+}\left( \gamma =0\right) $ is valid only when $G\leq G_{c}$
with

\begin{equation}
G_{c}=\frac{1}{2}\sqrt{\frac{\eta }{\Phi \left( \eta ,\varphi \right) }}.
\label{90-1}
\end{equation}

The extremum equation 
\begin{equation*}
\frac{\partial \varepsilon _{\pm }\left( \gamma \right) }{\partial \gamma }%
=0,
\end{equation*}%
does not have a non-zero solution for spin down state $\varepsilon _{-}$.
The non-zero solution for the spin-up state $\varepsilon _{+}$ is found as%
\begin{equation}
n_{p}=\frac{\gamma _{c}^{2}}{N}=\frac{1-\frac{2^{4}}{\eta ^{2}}G^{4}\Phi
^{2}\left( \eta ,\varphi \right) }{2^{3}G^{2}\Phi \left( \eta ,\varphi
\right) }  \label{91-1}
\end{equation}%
under the condition $G\leq G_{c}$. However \ the non-zero solution $%
\varepsilon _{+}\left( \gamma _{c}\right) $ is unstable since the
second-order derivative is negative $\frac{\partial ^{2}\varepsilon _{+}}{%
\partial \gamma ^{2}}\left( \gamma _{c}\right) <0$. The average energy of
unstable state $\varepsilon _{+}\left( \gamma _{c}\right) $ is given by%
\begin{equation}
\varepsilon _{+}\left( n_{p}\right) =\omega n_{p}+\Omega \sqrt{%
1-2^{3}G^{2}n_{p}\Phi \left( \eta ,\varphi \right) },  \label{95}
\end{equation}%
which is real for the valid photon number $n_{p}\geq 0$ in Eq. (\ref{91-1}).
Thus the complex energies and EP are excluded in the ground state. Fig. \ref%
{fig:9} displays the energy spectrum, photon number and atom energy $%
\varepsilon _{a}=\varepsilon _{+}-\omega n_{p}$ for the unstable state $%
\widetilde{S}_{+}$.

\begin{figure}[th]
\centering
\includegraphics[width=3in]{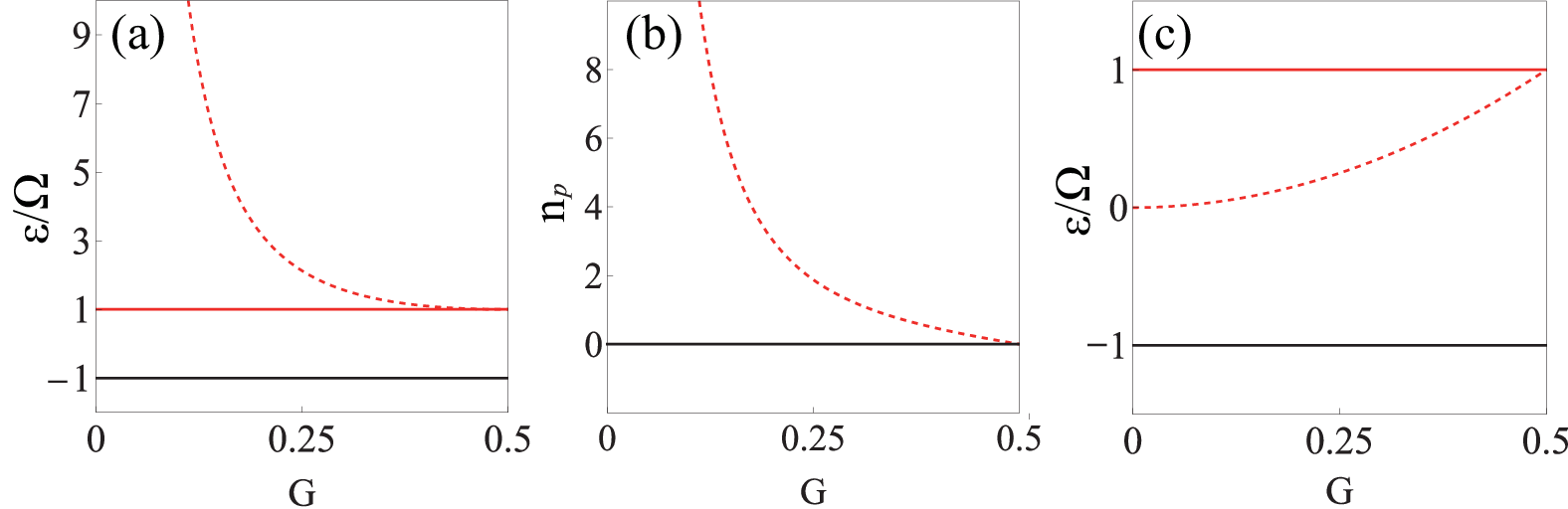}
\caption{ Average energy $\protect\varepsilon $ (a) photon number $n_{p}$
(b) and \ atom energy $\protect\varepsilon _{a}$ (c) as functions of the
coupling strength $G$ in the resonance $\protect\eta =1$, at field phase
angle $\protect\varphi =\protect\pi /3$. $\protect\widetilde{S}_{+}$ (red
dash curve) denotes the unstable superradiant state. $N_{\pm }$ (red, black
solid curves) denote the normal states.}
\label{fig:9}
\end{figure}

\begin{figure}[th]
\centering
\includegraphics[width=2.5in]{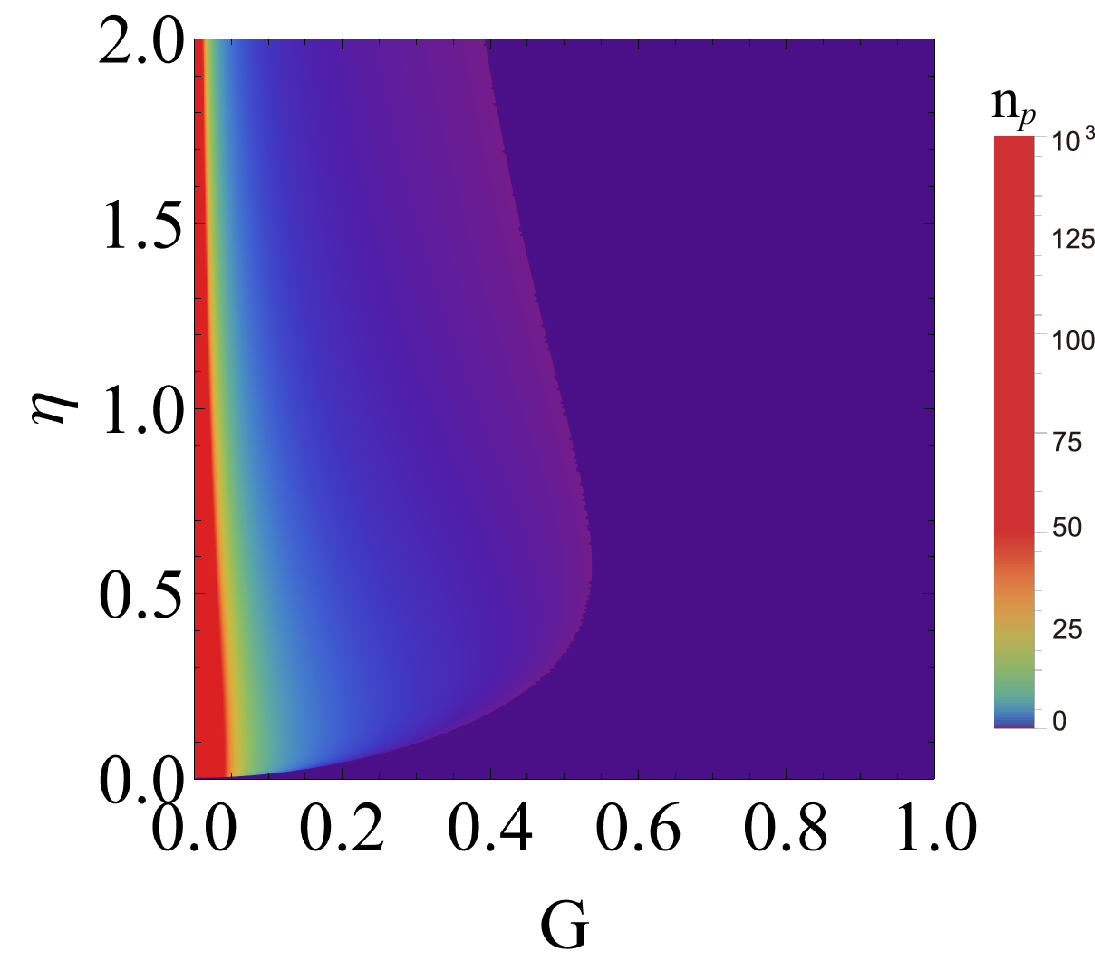}
\caption{$G-\protect\eta $ phase diagram in the unified gauge with the field
phase angle $\protect\varphi =\protect\pi /3$.}
\label{fig:10}
\end{figure}
Fig. \ref{fig:10} is the $G-\eta $ phase diagram in unified gauge, where the
photon number of state $\widetilde{S}_{+}$ is measured in color scale. $%
NP\left( N_{-}\right) $ denoting the NP with the normal state $N_{-}$ (dark
blue region) is located on the right of phase boundary $G_{c}$. $%
NP_{co}\left( N_{-},N_{+},\widetilde{S}_{+}\right) $ means the $NP\left(
N_{-}\right) $ coexisting with higher energy states $N_{+},\widetilde{S}_{+}$%
. The SP does not exists since superradiant state $\widetilde{S}_{+}$ is
unstable due to the photon number loss induced by the non-Hermitian
interaction. The QPT does not appear either.

\section{\textbf{CONCLUSION}}

We clarify in this article the gauge equivalence for the atom field
interaction. Since the Hamiltonian is explicitly time dependent with the
time varying field, the gauge transformation has to be applied on the
time-dependent Schr\H{o}dinger equation. Not only the transformed
Hamiltonian but also the basis vectors should be taken into account in the
new gauge. Thus the Hamiltonian called unified gauge is truly gauge
equivalent to the minimum coupling principle Eq. (\ref{2}). The Hamiltonians
of three gauges are compared with the system of three-level atom in an
optical cavity. The Hamiltonian of unified gauge includes both interactions
of Coulomb and dipole gauges.

We solve the DM Hamiltonians of $\Xi $-type three-level atoms in terms of
SCS variational method, where the three-level atom is described by
pseudospin operators of spin-$1$. The variational ground state of unified
gauge depends on the photon number as well as general phase-angle of cavity
field. The results of Coulomb and dipole gauges are just the special cases
of the unified gauge. A remarkable observation is that the ground states of
three gauges coincide exactly in the resonance condition $\eta =1$. The QPT
is analyzed explicitly as a comparison of three gauges. We reveal a
phase-sensitive effect in the unified gauge for the three-level system,
where photon-number distributions depend critically on the initial phase of
the optical field. This constitutes a crucial feature widely overlooked in
the two-level atom DM with the usual Coulomb gauge interaction \cite%
{PhysRevA.90.023622,PhysRevA.93.033630,Lian_2012,PhysRevA.110.063320}.

The coupling constant between cavity-field and atom is proportional to the
dipole matrix elements $\mathbf{x}_{ij}$, which are known as complex values.
The non-Hermitian Hamiltonian is obtained naturally by choosing the real
part of dipole matrix element. The semiclassical energy function averaged in
the trial wave function is indeed not always real. It becomes complex beyond
the EP $G_{ep}$. However the variational ground-state energy remains real
excluding the complex value as well as the EP. The superradiant state
becomes unstable because of photon number loss induced by the non-Hermitian
interaction. Thus the QPT does not exist in the non-Hermitian DM Hamiltonian.

\section*{ACKNOWLEDGMENTS}

This work was supported by National Natural Science Foundation of China
(Grants No. 12374312, No. 12074232, and No. 12125406), National Key Research
and Development Program of China (Grant No. 2022YFA1404500), and Shanxi
Scholarship Council of China (Grants No. 2022-014, No. 2023-033, and No.
2023-028).

\bibliographystyle{plain}
\bibliography{myref1}

\end{document}